\documentclass[journal]{IEEEtran}
\usepackage{authblk}

\usepackage{graphicx}
\graphicspath{ {images/} }
\usepackage{graphics}
\usepackage{algorithm}
\usepackage{amsmath,bm,amssymb}
\usepackage{algcompatible}
\usepackage{globalThings}
\usepackage{color}
\usepackage{fancyhdr}
\usepackage{setspace}
\usepackage{xfrac}
\usepackage{longtable}
\usepackage{import}

\ifCLASSINFOpdf
\else
\fi
\usepackage{lipsum}

\begin{document}
%

\title{Guided Generative Adversarial Neural Network for Representation Learning and High Fidelity Audio Generation using Fewer Labelled Audio Data}
%
%
%





\author[1]{Kazi Nazmul Haque}
\author[1]{Rajib Rana}
\author[2]{John H. L. Hansen}
\author[3,4]{Björn W Schuller}

\affil[1]{University of Southern Queensland, Australia}
\affil[2]{University of Texas at Dallas, USA}
\affil[3]{Imperial College London, UK}
\affil[4]{University of Augsburg, Germany}

\maketitle

\begin{abstract}

Recent improvements in Generative Adversarial Neural Networks (GANs) have shown their ability to generate higher quality samples as well as to learn good representations for transfer learning. Most of the representation learning methods based on GANs learn representations ignoring their post-use scenario, which can lead to increased generalisation ability. However, the model can become redundant if it is intended for a specific task. For example, assume we have a vast unlabelled audio dataset, and we want to learn a representation from this dataset so that it can be used to improve the emotion recognition performance of a small labelled audio dataset. During the representation learning training, if the model does not know the post emotion recognition task, it can completely ignore emotion-related characteristics in the learnt representation. This is a fundamental challenge for any unsupervised representation learning model. In this paper, we aim to address this challenge by proposing a novel GAN framework: Guided Generative Neural Network (GGAN), which guides a GAN to focus on learning desired representations and generating superior quality samples for audio data leveraging fewer labelled samples. Experimental results show that using a very small amount of labelled data as guidance, a GGAN learns significantly better representations.

\end{abstract}


%
\IEEEpeerreviewmaketitle

\section{Introduction}
Learning a meaningful representation from an unlabelled dataset is a challenging task \cite{goodfellow_book:2016}. Encouragingly, the advancement of the Generative Adversarial Neural Network (GAN) \cite{goodfellow:2014} concept has offered great success \cite{donahue2019large, dumoulin2016adversarially, DonahueKD16} in this field. In particular, the literature shows that the superior generation power of a GAN is useful for learning a good representation~\cite{karras:2017, karras2019style, Andrew_biggan, donahue2019large}. \RR{While the successes of GANs are mainly found in the image generation domain, they do not perform similarly well in the audio domain, because the audio structure is considerably more complicated than the one of an image and successful audio generation depends on generating different temporal scales accurately \cite{engel2019gansynth}, which makes the task harder for a GAN.}  Recently,  some studies \cite{engel2019gansynth,chris_wspecgan,andrTIFGAN} have shown intriguing results while using a GAN for audio generation. This has encouraged further studies of these methods to learn a representation from unlabelled audio data. 

Direct generation of the raw audio waveform is a very complex task. The wavenet generative neural network \cite{waveNet} achieved some success in such raw waveform generation; however, the proposed method was computationally expensive and slow. Rather than using GANs to generate a raw waveform, researchers are currently focusing on strategies to generate audio spectrograms and converting these spectrograms to audio \cite{GANSynth08710, chris_wspecgan,marafioti2019adversarial}. In this process, if the spectrogram can be generated successfully, the next challenge becomes the conversion of the spectrogram to the audio. The authors introducing TiFGAN  \cite{marafioti2019adversarial} use a phase-gradient heap integration (PGHI) \cite{zden_25} method to reconstruct audio from a spectrogram with a minimal loss.

Among the GAN architectures, the BigGAN model performs better in terms of image generation, and also in terms of representation learning \cite{Andrew_biggan}. However, to achieve a better result with a BigGAN architecture, we need to provide an enormous amount of data with labels \cite{lucic2019highfidelity}. Although it is easy to find according enormous open-access datasets, obtaining labels for these is very expensive. The recent work of DeepMind in the context of BigGANs has achieved state of the art (SOTA) generation \cite{lucic2019highfidelity} performance using fewer labels; however, these methods are not suitable for representation learning.


To address the problem of limited availability of labelled data, researchers are focusing on representation learning from unlabelled data. \RR{ In this paper, we are particularly interested in learning disentangled representations which offers the variations in any dataset to be readily separable in the representation space \cite{bengio2013representation,peters2017elements,bengio2013representation,goodfellow_book:2016, Michael05069}.} Although some GAN based studies have claimed good results for learning disentangled representation from unlabelled data \cite{radford2015,zhao:2017,chen2016infogan,makhzani:2016,chorowski_wavenet_autoencoder,karras2019style,donahue2019large}, recent work of Google AI shows both theoretically and practically that unsupervised disentangled representation learning is fundamentally impossible and some form of supervision is necessary to achieve this goal \cite{francesco_2019}. Therefore, for learning a good representation using GANs, there persists a requirement of good generation as well as of some form of supervision.
In this contribution, we address these challenges as follows. 


\begin{itemize}
\item We propose a new guided unsupervised representation learning model named Guided Generative Adversarial Neural Network (GGAN), which can guide the GAN model with a small labelled data sample to focus on specific characteristics when learning the representation as well as generating superior quality audio samples.

\item We evaluate the performance of the newly introduced GGAN applying the widely used Speech Command Dataset (S09) and Librispeech datasets. There, using gender information in Librispeech as a guide, we learn a representation from the S09 set and generate high-quality speech commands in male/female voices. A comparison with the existing studies shows that the novel GGAN performs significantly better than the state-of-the-art methods.



\end{itemize}

\section{Background and Related Work}
Among the successes in machine learning \cite{zhu2011heterogeneous, shao2014transfer, huh2016makes, shin2016deep, Rana7486123, latif2019direct, rana2019multi, li_PAMI,haque2018image}, the  Generative Adversarial Neural Networks (GANs) have recently brought extraordinary accomplishments in different research domains. Here the GANs are composed of two neural networks, a generator and a discriminator, which are trained based on a minimax game. During training, the discriminator tries to distinguish between real samples from the data distribution and fake samples generated from the generator, while the generator tries to fool the discriminator by producing samples closer to the real sample \cite{goodfellow:2014}. The generator maps any given random continuous sample distribution to the real data distribution. During this mapping, the generator learns significant latent characteristics of the data and tries to disentangle them in the random sample space \cite{radford2015}. 

In a supervised setting (e.\,g., a conditional GAN), the generator learns to map the random continuous distribution along with a  categorical distribution representing the labels of the dataset, to real samples from the dataset. Accurate labels of the dataset are fed to the discriminator to train these models. Although a GAN performs the best in a fully supervised setting, due to limited availability of labelled data, it is not possible to achieve fully supervised training in most practical scenarios. GANs such as  InfoGAN~\cite{chen2016infogan}, which are trained in a completely unsupervised way can avoid the need for labelled data.  In the case of the InfoGAN, true labels are not necessary. The generator can learn to map random categorical and continuous samples to real samples, maximising the information between the categorical distribution and generating samples with the help of another small network. However, this method fails to disentangle important features of the data when the complexity of the dataset is very high, such as images with higher resolution.

While fully supervised and fully unsupervised training both have their challenges, semi supervision \cite{zhu2005semi}---a middle-ground strategy---has received much interest in the past. Semi-supervised learning combines a small amount of labelled data with a large amount of unlabelled data during training. Guided unsupervised methods also fall into the category of semi-supervised learning. In~\cite{Spurr2017}, the authors propose a method to guide an InfoGAN. They use a small number of labels to help the InfoGAN to capture a specific representation. However, it fails to perform superior in cases like in complex datasets such as SVHN \cite{netzer2011reading}, CelebA\cite{liu2015faceattributes}, and CIFAR-10 \cite{krizhevsky2009learning}.  
In another work \cite{springenberg2015unsupervised}, the authors suggest learning a classifier from a partially labelled dataset and then use that classifier for semi-supervision in any GAN architecture. Kumar \cite{sricharan2017semisupervised} proposed two discriminators in a GAN architecture where one discriminator learns to identify real or fake samples from the unlabelled dataset, and another discriminator learns to identify real or fake samples with their labels from some labelled dataset. 
In a recent contribution of Google research, the authors explore different semi-supervised methods \cite{lucic2019highfidelity}. They predict the missing labels of the dataset with the help of a small labelled dataset. First, they train with self supervision, and then fine-tune the classifier with a small labelled dataset. Subsequently, they predict the labels for other missing labelled datasets. They also propose a co-training method for this task, \RR{where they train this classifier on top of the discriminator during the training.}

From the above, we identify two major gaps in the literature of semi-supervision using GANs. First, most of the proposed methods have shown promising results in term of generating higher quality samples from fewer labelled datasets; however, they do not offer any representation learning strategy.  Second, most of these latest studies are based on image datasets, and very few researchers have investigated the compatibility of these models in other domains such as audio.  
To contribute to this field, we propose the GGAN model which is capable of learning powerful representation from an unlabelled audio dataset with some guidance from a minimal amount of labelled samples. The newly introduced GGAN can as well generate higher quality audio samples at the same time.


\section{Proposed Research Methods}

\begin{figure*}[ht]
    \centering
    \frame{\includegraphics[width=.8\linewidth]{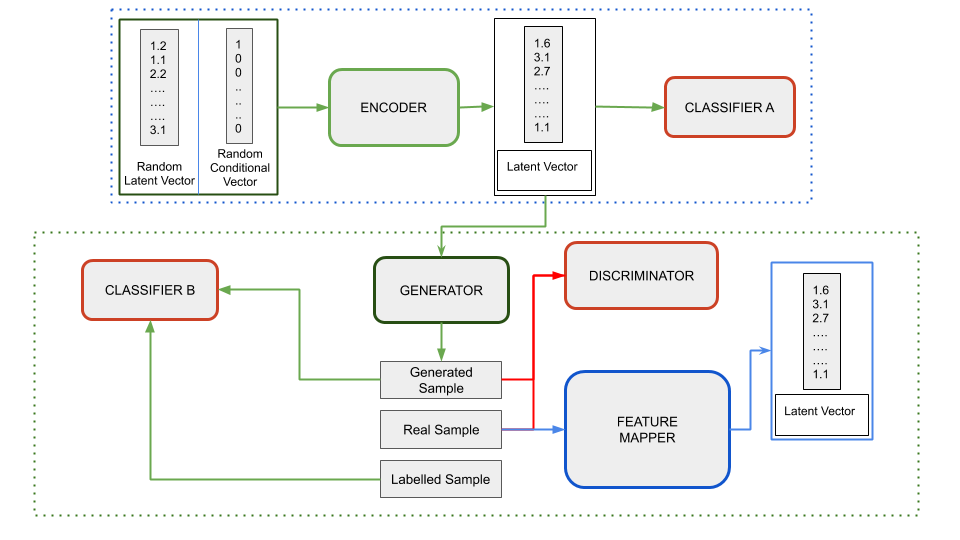}}
    \caption{ The connection between different networks named Encoder, Classifier A, Generator, Classifier B, Feature Extractor, and Discriminator is shown in this figure for an intuitive explanation of the working mechanisms of the proposed GGAN model.}
    \label{fig:mood1}
\end{figure*}

\subsection{Model Intuition}

A generator usually takes random latent vectors as input and generates samples from the real data distribution. In our proposed model, rather than feeding a random latent space to the generator, we feed it with a generated latent space, which is easily classifiable into $n$ categories. In the data distribution, the information of different categories are entangled and not easily separable, but in the latent space, these categories are disentangled, and we can separate them easily.
Our aim is that the generator learns to generate different data categories for different latent space categories. 
However, the challenge is how to force the generator to create such different classes of samples for different categories of latent space. To guide the generator, we hence use some labelled datasets and provide guidance from a classifier network. 



In Fig.~\ref{fig:mood1}, we show the essential connections between the parts of the model. In our model, we have an encoder that takes any random latent vector and random conditional vector to generate a new latent vector, which can be classified into $n$ categories through `classifier A'. The classifier network tries to predict the random conditional vector given the latent vector. This is achieved as the encoder, and the classifier network work together to maximise the information between the random categorical distribution (random conditional vector) and the generated latent distribution. 

The generator takes this new latent cector and generates samples from a real data distribution, and `classifier B' tries to classify this generated sample into $n$ categories. Through classifier B, it is ensured that for different given conditions in the encoder, the generator generates different categories of the sample. Now, the problem for the generator is to generate only the categories that are of our interest. In order that the generator can generate correct categories, we jointly train it with classifier B to classify  $n$ different categories of the dataset with the help of a few labelled data samples. Using the labelled data samples, classifier B gets trained to classify our concerned categories; it, therefore, classifies the generated samples into those target categories. As the encoder, the generator, and the classifier B are trained together, the generator tries to minimise the loss of classifier B by matching the classification label of classifier B given a latent vector.

 With the help of classifier B, the generator thus learns to create $n$ numbers of categories of the sample from $n$ different categories of latent space generated by the encoder. As classifier B is trained with the small labelled dataset,  it will be able to classify some categories correctly from the dataset at the start of the training. As the training goes on, the generator will generate other similar samples for each of the categories, which will improve the classification accuracy of classifier B. 
 

In the end, the generator is trained to generate $n$ categories samples from $n$ categories in the latent space generated by the encoder. In the latent space, the categories of the dataset are disentangled and easily classifiable by the classifier network, which makes latent space a powerful representation of the data distribution. We, therefore, connect a feature extractor network into this framework which learns to map real samples to the latent space. The output of the feature extractor then essentially becomes our ``learned representation''.

\begin{figure}[t!]
\centering
    \frame{\includegraphics[width=1\columnwidth]{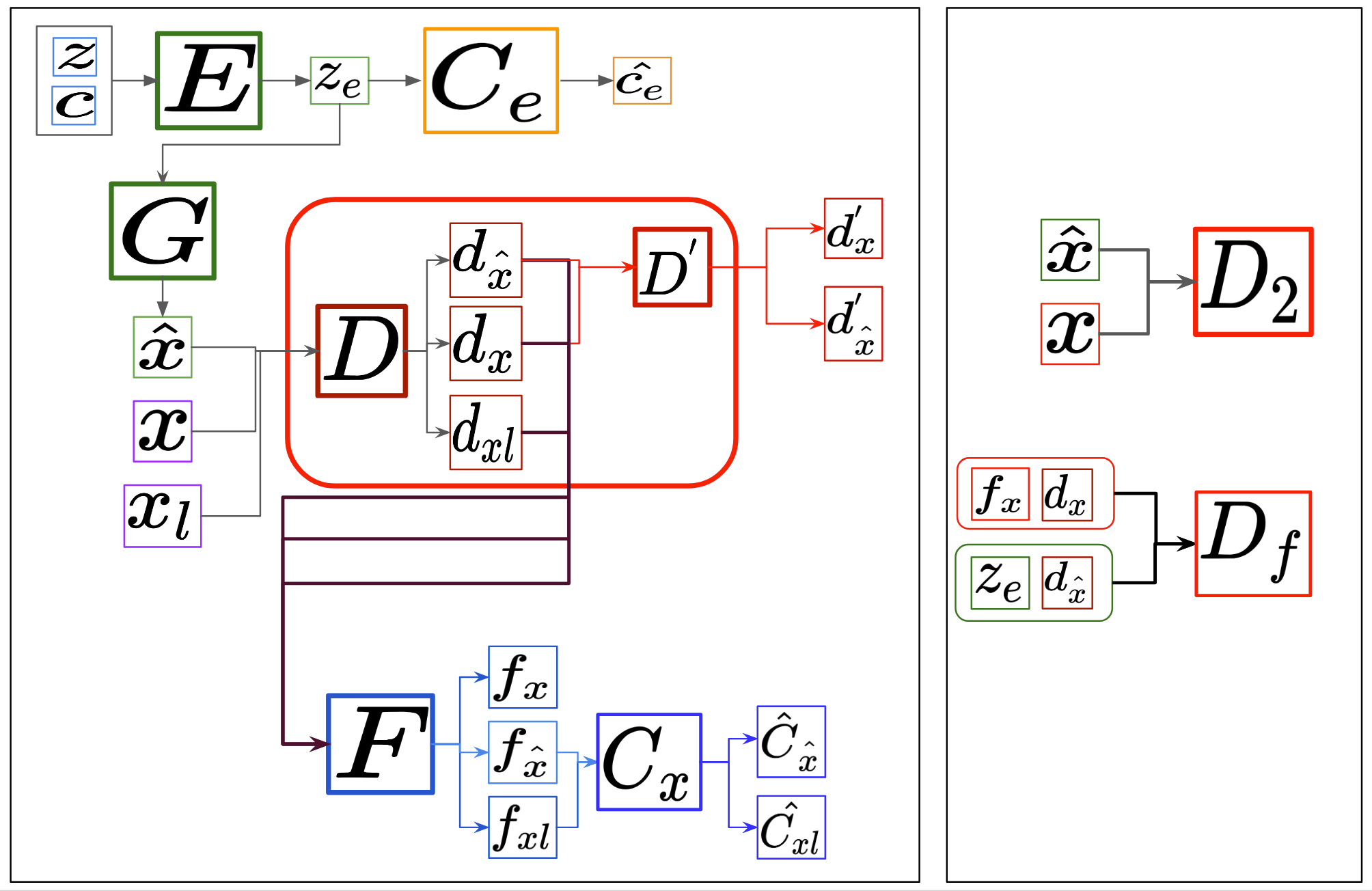}}
    \caption{ The architecture of the proposed GGAN model. The model consists of the Encoder $E$, Generator $G$, Feature Extractor $F$, two Classifiers $C_{e}$ and $C_{x}$, three Discriminators $D_1$,$D_2$, and $D_{f}$ Networks. }
    \label{fig:mood2}
\end{figure}

\subsection{Architecture of the GGAN}

In our proposed model, we have eight networks. These are the Encoder $E$, Generator $G$, Feature Extractor $F$, two Classifiers $C_{e}$ and $C_{x}$, and three Discriminators $D_1$,$D_2$, and $D_{f}$. The description is as follows. 

\subsubsection{Encoder and Classifier ($C_e$)}
The Encoder, $E$ learns to map any sample $z$ and $c$ to $z_{e}$ $\in$ $u(z)$, where $u(z)$ is any continuous distribution generated by $E$, $z$ $\in$ $p(z)$, and $c$ $\in$ $Cat(c)$ ; $p_z$ is a random continuous distribution, e.\,g., a continuous uniform distribution and $Cat(c)$ is a random categorical distribution.

When $E$ learns to map $z$ and $c$ to $z_{e}$, it can easily ignore the categorical distribution. To address this problem, we introduce a classifier network $C_{e}$, which takes $z_{e}$ as input and outputs the predicted class $\hat{c_{e}}$, where the true label is the given categorical sample $c$. By using the network $C_{e}$, we force $E$ to maximise the mutual information between $c$ and $z_{e}$. Suppose we have $n$ categories in the  $Cat(c)$, then $E$ learns to create a sample space $u(z)$, which can be classified into $n$ categories by the network $C_{e}$.

\subsubsection{Generator}
 Like any other GAN framework, we have a Generator $G$, which learns to map $z_{e}$ into sample $\hat{x}$ $\in$ $P_{G}$, where $P_{G}$ is the generated sample distribution. One of the major goals of $G$ is to generate $P_{G}$ so that it matches the true data distribution $P_{data}$. Another goal of $G$ is to maximise the mutual information between $\hat{x}$ and the given random condition $c$ (random categorical sample) ensuring there are categorical variations in the generated samples. 

\subsubsection{First ($D_{1}$) and Second ($D_{2}$) Discriminators }
 The Discriminator $D_1$ has two parts: a feature extraction part, $D$, and the real/fake sample identification part $D^{'}$. From $D$ we obtain the features $d_{\hat{x}}$, $d_{x}$, and $d_{x_l}$ for $\hat{x}$, $x$, $x_l$, respectively, where a real data sample is $x$ $\in$ $P_{data}$ and a labelled data sample resembles $x_l$ $\in$ $P_{ldata}$ ; $P_{ldata}$ is the labelled data distribution. Here,   $P_{ldata}$ can be the subset of the $P_{data}$ or any other data distribution. For a real sample $x$  and a fake sample $\hat{x}$, we obtain the output respectively as $d^{'}_{x}$ and $d^{'}_{\hat{x}}$, from $D^{'}$. 
In our $D_1$ Network, $D$ optimises the feature learning for both the classification and the discrimination tasks, which renders the optimisation task considerably complex. So, to avoid further complicating of the optimisation task, we use another Discriminator $D_{2}$ to merely perform a real and fake sample discrimination.

\subsubsection{Feature Extractor, Classifier ($C_x$), and Third Discriminator ($D_{f}$)}

 The feature extractor network $F$ learns the representation  $f_{\hat{x}}$ , $f_{x}$, and $f_{xl}$ for $\hat{x}$, $x$ and $xl$, respectively.
Rather than feeding directly samples to $F$, we feed the features ($d_{\hat{x}}$, $d_{x}$ and to $d_{x_l}$) of these samples from $D$. This is to make it computationally inexpensive, also to reduce the chance of overfitting as the features from $D$ are constantly changing. 

$F$ is trained to map $\hat{x}$, $x$ and $xl$ to the latent space $u(z)$.
To ensure this, we have another discriminator $D_{f}$ which learns to identify an $(f_{x},x)$ pair as a fake, and a $(z_{e},\hat{x})$ pair as a real pair. Similarly, rather than feeding directly $x$ and $\hat{x}$, we feed their features from $D$. 

The second Classifier $C_x$ learns to classify labelled data as well as it learns to classify the generated image $\hat{x}$ according to a given categorical, conditional sample $c$. For the classification of $\hat{x}$ and $xl$, we feed the features generated from $F$ to $C_x$.

\subsection{Losses}


\subsubsection{Encoder, Classifier ($C_{e}$), and Generator loss}

 For the Encoder $E$ and the Classifier $C_{e}$, let the classification loss be $EC_{loss}$. We have $z_{e} = E(z,c)$, therefore,

\begin{equation}
\label{eq:1}
\begin{aligned}
EC_{loss} = - \sum c \log (C_{e}(z_{e})).\\
\end{aligned}
\end{equation}

 For the Generator $G$, we have two generation losses coming from the discriminator $D_1$ and $D_{2}$. For the generator and discriminator, we use the hinge loss. We have $\hat{x} = G(z_{e})$ and $d_{\hat{x}} = D(\hat{x})$, therefore,

\begin{equation}
\label{eq:2}
\begin{aligned}
G_{loss1} = -D^{'}(d_{\hat{x}}), \\
\end{aligned}
\end{equation}
\begin{equation}
\label{eq:2-1}
\begin{aligned}
G_{loss2} = -D_{2}(\hat{x}). \\
\end{aligned}
\end{equation}

 The Generator $G$ has another loss for the classification of the sample. We have $f_{\hat{x}} = F(d_{\hat{x}})$, so the Classification Loss, $GC_{loss}$ for $G$ is,

\begin{equation}
\label{eq:3}
\begin{aligned}
GC_{loss} = - \sum c \log (C_{x}(f_{\hat{x}}).\\
\end{aligned}
\end{equation}

For mode collapse, we define a loss named ``Mode Divergence Loss'', $MG_{loss}$. For calculating this loss, we take two random inputs $z_1$ and $z_2$, and the same conditional code $c$. For $z_1, z_2$, we get $\hat{x_{1}}$ = $G(E(z1,c))$ and $\hat{x_{2}}$ = $G(E(z2,c))$, respectively. 

We also take two random samples $x_{1}$ and $x_{2}$ from the real data distribution $p_{data}$. We calculate the loss based on the feature extracted from $D$. Let $d_{x1}$ = $D(x_{1})$,  $d_{x_{2}}$ = $D(x2)$, $d_{\hat{x1}}$ = $D(\hat{x_{1}})$, $d_{\hat{x2}}$ = $D(\hat{x_{2}})$, and $\alpha$ is a small parameter like .0001, so we get,

\begin{equation}
\label{eq:4}
\begin{aligned}
MG_{loss} = \max \{1,\sfrac{\sum{(|d_{x1} - d_{x2}|)}}({\sum{(|d_{\hat{x1}} - d_{\hat{x2}}|)}} + \alpha) \}.
\end{aligned}
\end{equation}

\noindent Hence, we have a combined loss, $ECG_{loss}$ for $E$, $C$, and $G$. We averaged the $G_{loss1}$, $G_{loss2}$, and $MG_{loss}$ as all of these are losses for the generation of the sample. The $E$, $C$, and $G$ networks are updated to minimise the loss $ECG_{loss}$.

\begin{equation}
\label{eq:5}
\begin{aligned}
ECG_{loss} = (G_{loss1} + G_{loss2} + MG_{loss})/3 \\+ EC_{loss} + GC_{loss}.
\end{aligned}
\end{equation}

\subsubsection{Feature Extractor and Classifier ($C_x$) loss}

We have the feature generation loss, $FG_{loss}$, coming from the third Discriminator $D_{f}$, so that $F$ creates features like $z_{e}$ from real data.

\begin{equation}
\label{eq:6}
\begin{aligned}
FG_{loss} = -D_{f}(f_{x},d_{x}). \\
\end{aligned}
\end{equation}
We have two classification losses for $C_{x}$, one for the labelled sample, $Cl_{loss}$, and another one for the generated sample, $Cg_{loss}$. If the label of the real sample is $y$, and we have $f_{xl} = F(d_{xl})$ and $f_{\hat{x}} = F(d_{\hat{x}})$, therefore,
\begin{equation}
\label{eq:6}
\begin{aligned}
Cl_{loss} = - \sum y \log (C_{x}(f_{xl})),\\
\end{aligned}
\end{equation}
\begin{equation}
\label{eq:6-1}
\begin{aligned}
Cg_{loss} = - \sum c \log (C_{x}(f_{\hat{x}})).\\
\end{aligned}
\end{equation}
Likewise, the total loss for $F$ and $C_{x}$ is $FC_{loss}$, which is the sum of the above losses: 
\begin{equation}
\label{eq:6}
\begin{aligned}
FC_{loss} = Cl_{loss} + Cg_{loss} + FG_{loss}.
\end{aligned}
\end{equation}
We update the $F$ and $C_{x}$ network to minimise the $FC_{loss}$.

\subsubsection{Discriminators loss}
The $D^{'}$ part of $D_{1}$, and $D_{2}$ are two discriminators for identifying the real/fake samples generated from the generator. $D_1$ also has a part $D$, which is responsible for generating features for the $F$ and $C_{x}$ networks. It is also optimised to reduce the classification loss, $Cl_{loss}$, of the real labelled samples. Finally, the Discriminator $D_{f}$ identifies the real or fake feature sample pairs from $F$. The discriminator losses $D_{1loss}$, $D_{2loss}$ and $D_{floss}$ for $D_1$, $D_2$, and $D_{f}$, respectively, are given by,
\begin{equation}
\label{eq:7}
\begin{aligned}
D_{1loss} = - min(0, -1+D^{'}(D(x))) \\- min(0,-1-D^{'}(D(\hat{x}))) - Cl_{loss}, \\
\end{aligned}
\end{equation}
\begin{equation}
\label{eq:8}
\begin{aligned}
D_{2loss} = - min(0, -1+D_{2}(x)) \\- min(0,-1-D_{2}(\hat{x})), \\
\end{aligned}
\end{equation}
\begin{equation}
\label{eq:9}
\begin{aligned}
D_{floss} = - min(0, -1-D_{f}(f_{x},d_{x})) \\- min(0,-1+D_{f}(z_{e}, d_{\hat{x}}))). \\
\end{aligned}
\end{equation}
\noindent The discriminators' weights are updated to maximise these losses. The algorithm to train the whole model is given in Algorithm 1.

\begin{algorithm}[t!]
\caption{\small Minibatch stochastic gradient descent training of the proposed GGAN.
The hyperparameter $k$ represents the number of times the discriminators are updated in one iteration. We used $k=2$, which helped to converge faster.
}

\begin{algorithmic}[1]
\label{alg:AGF2}
\FOR{number of training iterations}
  \FOR{$k$ steps}
    
    \STATE {Sample minibatch of $m$, noise samples $ \{ \bm{z^{(1)}}, \dots, \bm{z^{(2m)}} \} $ from $p_z$, conditions $ \{ \bm{c^{(1)}}, \dots, \bm{c^{(m)}} \} $ from Cat(c),
    data points $ \{ \bm{x^{(1)}}, \dots, \bm{x^{(2m)}} \} $ from  $p_{data}$ and labelled data points $ \{ \bm{xl^{(1)}}, \dots, \bm{xl^{(m)}} \} $ from  $P_{ldata}$.}
    
    \STATE {Update the parts $D, D^{'}$ of discriminator $D_1$ by ascending its stochastic gradient:
        \[
            \nabla_{\theta_d, \theta_{d^{'}}} \frac{1}{m} \sum_{i=1}^m \left[\bm {{D_{1loss}}}^{(i)} \right].
        \]}

    \STATE {Update the discriminator $D_{2}$ by ascending its stochastic gradient:
        \[
            \nabla_{\theta_{d_{2}}} \frac{1}{m} \sum_{i=1}^m \left[\bm{{D_{2loss}}}^{(i)}\right].
        \]}   
        
    \STATE {Update the discriminator $D_{f}$ by ascending its stochastic gradient:
        \[
            \nabla_{\theta_{d_f}} \frac{1}{m} \sum_{i=1}^m \left[\bm{{D_{floss}}}^{(i)}\right].
        \]}

  \ENDFOR
   
    \STATE {Repeat step [3].}
    
    \STATE {Update the Generator $G$, Encoder $E$ and Classifier $C_{e}$ by descending its stochastic gradient:
        \[
            \nabla_{\theta_{g},\theta_{e},\theta_{c_e}} \frac{1}{m} \sum_{i=1}^m \left[\bm{{ECG_{loss}}}^{(i)} \right].
        \]}

    \STATE {Repeat step [3].}
    
    \STATE {Update the Feature Extractor $F$ and Classifier $C_{x}$ by descending its stochastic gradient:
        \[
            \nabla_{\theta_{f},\theta_{c_x}} \frac{1}{m} \sum_{i=1}^m \left[\bm{{FC_{loss}}}^{(i)} \right].
        \]}
     \ENDFOR

\end{algorithmic}
\end{algorithm}

\section{Data and Implementation Detail}
\subsection{Datasets}
For the validation of the GGAN, we use S09 \cite{Pete_03209} and the Librispeech dataset \cite{panayotov2015librispeech}. In the S09 dataset digits from zero to nine are uttered by 2618 speakers \cite{Pete_03209}. Most of the recent studies use the S09 dataset for evaluating their model for audio generation. This dataset is noisy and includes 23,000 samples where many of these samples are poorly labelled. Labels for the speakers and gender are not available in S09.

LibriSpeech is a corpus of approximately 1000 hours of 16kHz read English speech. In the dataset, there are 1166 speakers where 564 are female, and 602 are male.

\subsection{Measurement Metrics}
For evaluating generation we use Inception Score (IS)\cite{salimans:2016} and Fréchet Inception Distance (FID) \cite{heusel2017gans,barratt2018note}.

\subsubsection{Inception Score (IS)}
IS score measures the quality of the generated samples as well as the diversity in the generated sample distribution. Pretrained Inception Network V3 \cite{szegedy2014going} is used to get the labels for the generated samples. The conditional label distribution $p(y | x)$ is derived from inception network, where $x$ is the generated sample. We want the entropy to be low, which indicates the good quality of the image. It is also expected that the images are diverse,  so the marginal label distribution $ \int p(y | x) dz $ should have high entropy. Combining these two requirements, the KL-divergence between the conditional label distribution and the marginal label distribution is computed from metric $\exp(\mathbb{E}_{x} {KL}(p(y | x) || p(y)))$. As an inception network is used, it is called Inception Score (IS score). A higher IS score indicates the good quality of the generated samples.

\subsubsection{Fréchet Inception Distance (FID)}
IS score is computed solely on the generated samples. Fréchet Inception Distance (FID) improves the IS score by comparing the statistics of generated samples to real data samples. First the features are extracted for both real and generated samples from the intermediate layer of the inception network. Then the mean $\mu_{r}$, $ \mu_{g}$ and covariance  $\Sigma_{r}$, $\Sigma_{g}$ for real and generated samples are calculated respectively from those features. Finally, the Fréchet Distance \cite{dowson1982frechet} between two multivariate Gaussian distributions (given by the $\mu_{r}$, $ \mu_{g}$ and $\Sigma_{r}$, $\Sigma_{g}$) is calculated using: $||\mu_r - \mu_g||^2 + \text{Tr} (\Sigma_r + \Sigma_g - 2 (\Sigma_r \Sigma_g)^{1/2})$. A lower FID score indicates the good quality of the generated samples.


\RR{Note that the inception v3 model is trained on imagenet  dataset~\cite{imagenet_cvpr09}, which is completely different from audio spectrograms. Therefore, it will not be able to classify spectrograms into any meaningful categories, resulting in poor performance on the calculation of IS and FID score for our datasets.}
In the "Adversarial Audio Synthesis"\cite{chris_wspecgan} paper, instead of using the pretrained inception v3 network to calculate the IS and FID scores, the authors train a classifier network on S09 spoken digit dataset and obtain good performance. We, therefore, use their pretrained model to calculate both IS and FID scores for our generated samples.

\subsection{Experimental Setup}
\label{sub:Experimental_Design}

Performance of GGAN model is evaluated based on two tasks, Representation Learning and Audio Generation. To evaluate the performance of audio generation, we compare IS and FID score of GGAN with that of supervised and unsupervised BigGAN as well as related research works on S09 dataset. First, the generated log-magnitude spectrograms of the models are converted to audio with the PGHI algorithm and then we convert the audio back to spectrogram to pass through the pretrained classifier to calculate the IS and FID score as the input spectrogram format for the pretrained model is different. Therefore, essentially the evaluation is done based on the generated audio rather than evaluating on the spectrogram directly. 

For representation learning, we compare the GGAN model with unsupervised BigGAN and supervised Convolutional Neural Network (CNN).  Our primary goal is to learn representation from unlabelled S09 training dataset so that we can get better classification result on the S09 test dataset. For any GAN model, the latent space captures the representation of the training dataset so to map the real data samples to the latent space we follow the strategy from the 'Adversarially learned inference' paper\cite{dumoulin2016adversarially}. 

After training the unsupervised BigGAN, we train a feature extraction network to reverse map the sample to latent distribution to get the representation for real samples. Then we train a classifier at the top of the feature extraction network with 1 to 5\% randomly sampled labelled dataset from the training dataset and evaluate on the test dataset. A sampling of the training dataset was repeated five times, and the results were averaged. We use the same 1 to 5\% labelled data for the training of a CNN network. For this setting, we use the same CNN architecture as feature extraction network. Also, these 1 to 5\% labelled data
was used during the training of the GGAN model as guidance. 


We take the pretrained $D$, $F$ and $C_{x}$ networks from GGAN models, pass the test dataset through those pretrained networks to get the prediction for classes $C_{x}(F(D(x_{test})))$. Then we compute the classification accuracy on the test dataset. With these experiments, we can evaluate the effectiveness of using a small labelled training data during or after the training of the model. So these experiments will demonstrate if the guidance in GGAN, can improve the performance. 

To evaluate the quality of the representation learnt we visualise the representation space. We then evaluate the disentanglement by observing the linear interpolation in the representation space.

We also test the possibility of guiding an unsupervised model to learn an attribute of data in a dataset, where the guidance comes from a non-related dataset. Here we have used the whole S09 training data as the unlabelled dataset and Librispeech as the labelled dataset for guidance on the gender. Note that Gender information is not available for the S09 dataset. We have taken 50 male and 50 female speakers from the Librispeech dataset with 5 minutes of audio sample for each speaker. Our expected output from the GGAN is to produce the male and female spoken digits based on the guidance from the Librispeech dataset.

For our generator and discriminator network in the GGAN model, we use the BigGAN \cite{Andrew_biggan} architecture. We maintain the parameters the same as BigGAN, except we only change the input of the Discriminator and output of the Generator to accommodate the 128 by 256 size log-magnitude spectrogram. During the training of GGAN, we keep the learning rate of the generator and Discriminator equal, which boosts the IS score by 1. The architecture details are given in the supplementary document.






\section{Results}
\begin{figure*}[t!]
    \centering
    \frame{\includegraphics[width=\textwidth]{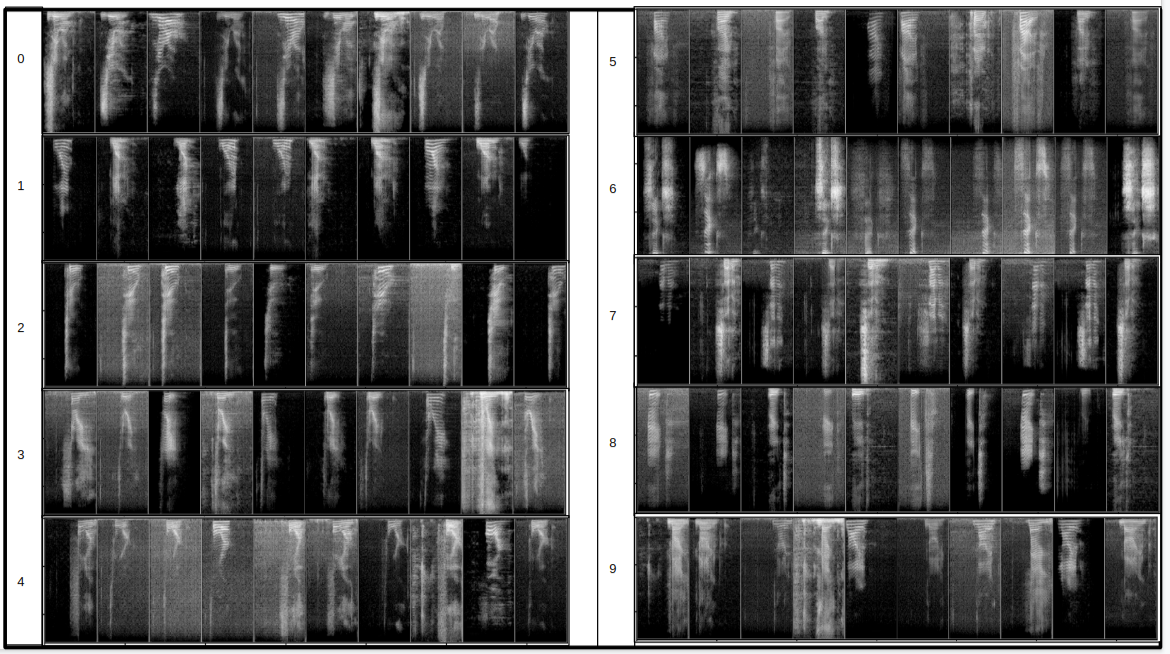}}
    \caption{ The generated log-magnitude spectrograms using the proposed GGAN model for different digit categories is shown in this figure. Each row shows a distinct digit category from 0 to 9,  where the columns show the diversity in the generated samples.  Numbers on the left side of each row show the digit category for that particular row.}
    \label{fig:fig3}
\end{figure*}

\begin{table}[t!]
\centering
\caption{Comparison between the performance of the GGAN model and the other models on the S09 dataset, in terms of the quality of the generated samples, measured with IS and FID score.}
\label{tab:my-table}
\begin{tabular}{|l|l|l|}\hline
\textbf{Model Name} & \textbf{IS Score} & \textbf{FID Score} \\ \hline
Real (Train Data) \cite{chris_wspecgan} & 9.18 $\pm$ 0.04 & - \\ \hline
Real (Test Data) \cite{chris_wspecgan} & 8.01 $\pm$ 0.24 &  -\\ \hline
TiFGAN \cite{andrTIFGAN} & 5.97 & 26.7 \\ \hline
WaveGAN \cite{chris_wspecgan} & 4.67 $\pm$ 0.01 &  -\\ \hline
SpecGAN \cite{chris_wspecgan} & 6.03 $\pm$ 0.04 &  -\\ \hline
Supervised BigGAN & 7.33 $\pm$ .01 & 24.40  $\pm$ 0.5  \\ \hline
Unsupervised BigGAN & 6.17 $\pm$ 0.2 & 24.72  $\pm$ 0.05 \\ \hline
{\bf GGAN} & {\bf $7.24 \pm 0.05$} & {\bf $25.75 \pm 0.1$} \\ \hline
\end{tabular}
\end{table}


For the unsupervised BigGAN model, we achieve an IS score of $6.17\pm 0.2$, and an FID score of $24.7$2, whereas we receive an IS score of $7.3 \pm 0.01$, and FID score of $24.40 \pm 0.5$ for the supervised BigGAN. For our GGAN model, with 5 per cent labelled data as guidance, we achieve a maximum IS score of $7.24\pm0.05$, and an FID score of $25.75 \pm 0.1$ for the generated samples, which is very close to the performance of the fully supervised BigGAN model and considerably better than unsupervised BigGAN model. Comparison of the IS score and FID score between different models are shown in the  table \ref{tab:my-table}. We observe that the performance of the proposed GGAN is better than that of other studies reported in the table \ref{tab:my-table}. Generated samples for different digit categories are shown in Figure \ref{fig:fig3}. Notice the diversity in the sample generation for any particular digit category.

A comparison on the test accuracy between the unsupervised BigGAN and GGAN is shown in table \ref{tab:my-table2}. With a 5\,\% labelled dataset, we have achieved an accuracy of $92 \pm 0.87$\,\%, which is close to the accuracy of the fully supervised CNN model ($96.2$\,\%). Therefore, guidance during training helps to learn better representation of the dataset.  

As we have guided the proposed GGAN model with the digit categories, it is expected that in the representation space, the digit categories are disentangled.
To investigate this scenario, we visualise the learnt representation in the 2D plain. We take the representation/feature $F(D(x_{test}))$ of the test dataset passing through the trained $D$ and $F$ networks. Then, the higher dimensional features are visualised with the t-SNE (t-distributed stochastic neighbour embedding) \cite{maaten2008visualizing} visualisation technique and shown in figure \ref{fig:fig4}. In the figure, we can observe that the features of the similar categories are clustered together and they are easily separable. 

\begin{figure*}[t!]
    \centering
    \frame{\includegraphics[width=\textwidth]{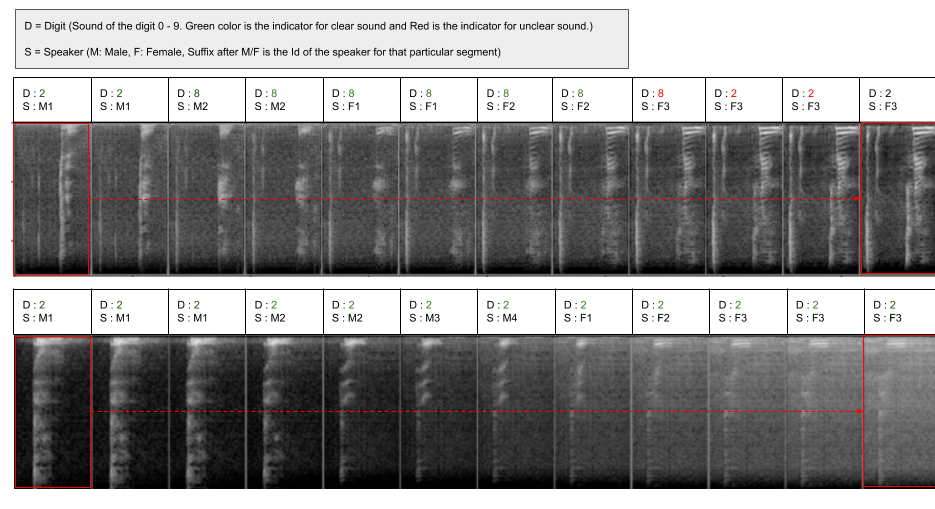}}
    \caption{ Generated spectrogram from the interpolation of the latent space from the male sound of digit 2 to the female sound of digit 2 (left to right). The top row is the linear interpolation for the unsupervised BigGAN, and the bottom row represents the linear interpolation for the proposed GGAN.}
    \label{fig:fig1}
\end{figure*}

To investigate the quality of the learnt latent space by the generator, we conduct a linear interpolation between two random points in the latent space as in the DCGAN work \cite{radford2015} and observe a smooth transition in the generated spectrogram space as well as that after converting them to audio, the transition is smooth. This indicates good learning of latent space by the generator. But for the unsupervised BigGAN, during the linear interpolation, we notice a smooth transition in the generated spectrogram space but after converting them to the audio, we find that the transaction becomes non-smooth. This indicates that the generator is able to capture the semantics of the spectrogram in the latent space but is unsuccessful to capture the semantics in the audio space. As the unsupervised GAN is trained with spectrograms, it has no idea about the characteristics of the audio samples. But with some guidance, we have shown that the generator of the GGAN can learn important semantics of audio though it is trained with spectrograms. 

In figure \ref{fig:fig1}, we show the linear interpolation for both the unsupervised BigGAN and the proposed GGAN. We avoid the latent space interpolation for the supervised BigGAN, as the generator of the supervised BigGAN uses both conditions, $c$ and latent space, $z$, during the sample generation, $G(z,c)$, which discourages the generator to learn any conditional characteristics in the latent space. It instead learns the common attributes in the latent space. In our S09 dataset experiment, the supervised BigGAN does not disentangle the digit categories (condition) in the latent space. It learns the common characteristics like gender, pitch, volume, noise, etc.\ in the latent space and generates different digits for the same latent space given different conditions.

\begin{table}[t!]
\centering
\caption{Comparison of the test data classification accuracy between a CNN, unsupervised GAN, and the proposed GGAN on the S09 dataset}
\label{tab:my-table2}
\begin{tabular}{|l|l|l|l|}\hline
\textbf{\begin{tabular}[c]{@{}l@{}}Training\\ Data Size\end{tabular}} &
\textbf{\begin{tabular}[c]{@{}l@{}}CNN\\ Network\end{tabular}} & \textbf{\begin{tabular}[c]{@{}l@{}}Unsupervised\\ GAN\end{tabular}} & \textbf{GGAN} \\ \hline
1\%& 82 $\pm$  1.2 & 73 $\pm$  1.02 &  84 $\pm$  2.24\\ \hline
2\% & 83  $\pm$ 0.34 & 75 $\pm$  0.41 &  85 $\pm$  1.24\\ \hline
3\% & 83  $\pm$ 0.23 & 78 $\pm$  0.07 & 88 $\pm$ 0.1 \\ \hline
4\% & 84  $\pm$ 0.34 & 80 $\pm$  0.01 &  91 $\pm$  0.5 \\ \hline
5\% & 84 $\pm$  1.02 & 80 $\pm$  1.72 &  92  $\pm$ 0.87\\ \hline
100\% & 96.2 & - &  - \\ \hline
\end{tabular}
\end{table}

\begin{figure}[t!]
    \centering
    \frame{\includegraphics[width=.8\linewidth]{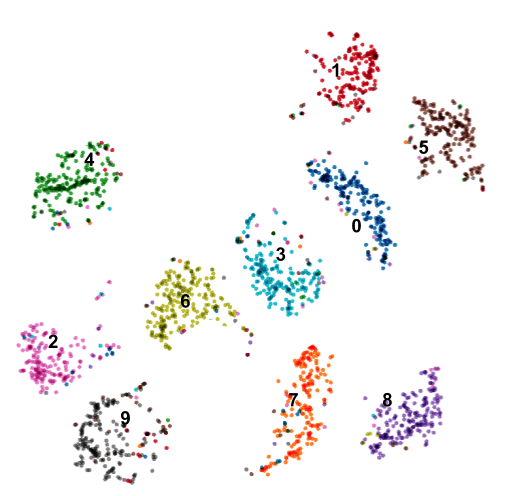}}
    \caption{ t-SNE visualisation of the learnt representation of the test data of the S09 dataset. Here, different colours of points represent different digit categories. Representations of the different digit categories are clustered together.}
    \label{fig:fig4}
\end{figure}


We also investigate the effectiveness of the guidance from training data of a completely separate dataset. As unlabelled dataset, we use the S09 dataset and for the labelled dataset part, we use the librispeech dataset with the labels for speaker gender (male and female). We successfully generate samples of 0-9 digits for males and female speakers. We achieve an IS score of $6.52 \pm 0.5$, and an FID score of $26.21 \pm 0.1$. Here, although our GGAN model is not guided for digit categories, it shows state of the art performance.


\section{Conclusion and Lesson Learnt}
In this contribution, we proposed the novel Guided Generative Adversarial Neural network (GGAN), where we guide an unsupervised GAN network with some labelled data. This allows the network to learn specific attributes while learning the representation of the dataset, which is useful for transfer learning or any other machine learning tasks. As we guide the model during training according to a post-task, the proposed GGAN can be used as a task-aware network that learns task-specific representation. We showed that the GGAN can learn powerful representations as well as can generate high-quality samples given some labelled data as guidance. 

Our model was successfully evaluated based on one-second audio data; it remains a challenge to make it useful for large size audio with higher complexity. However, any long audio can be divided into one-second chunks, and this model can be beneficial for generating high-quality chunks and learn useful representation which can be used for challenging tasks. While more research is necessary to understand this model, our research is likely to motivate others to work on the guided representation learning model where, with some guidance, one model can learn powerful representations from the enormous freely available unlabelled data. 

Among many minor challenges, the main challenge we face is related to the sample generation of the model. There was a severe issue of mode collapse, which we could not satisfyingly fix implementing different techniques in the literature \cite{salimans:2016}. However, we received  inspiration from the Mode Seeking GAN \cite{Qi05628}, modified the loss function, and created a unique feature loss, which immediately fixed the mode collapse problem. The feature loss calculates the ratio of the difference between the two real samples and two generated samples. If the generator creates very similar or the same samples, it gets penalised and tries to find more modes. \RR{Audio samples discussed in this paper, can be found in the GitHub link \underline{https://github.com/KnHuq/GGAN}.} 



\newpage
\medskip
\bibliographystyle{IEEEtran}
\bibliography{references}


%

\appendices

\newpage
\section{Supplement Material}

In this section, we discuss configurations of different architectures used in the paper. List of abbreviations used in the description of the architectures are shown in table \ref{tab:tab0} \cite{lucic2019highfidelity}.

\begin{table}[h!]
\centering
\caption{Abbreviations for defining the architectures.}
\begin{tabular}{|l|l|}
\hline
\textbf{\begin{tabular}[c]{@{}l@{}} Full Name\end{tabular}} & \textbf{\begin{tabular}[c]{@{}l@{}}Abbreviation\\ \end{tabular}} \\ \hline
Resample & RS \\ \hline
Batch normalisation & BN \\ \hline
Conditional batch normalisation & cBN \\ \hline
Downscale & D \\ \hline
Upscale & U \\ \hline
Spectral normalisation & SN \\ \hline
Input height & h \\ \hline
Input width & w \\ \hline
True label & y \\ \hline
Input channels & ci \\ \hline
Output channels & co \\ \hline
Number of channels  & ch \\ \hline
\end{tabular}
\label{tab:tab0}
\end{table}

\subsection{Supervised BigGAN}

\RR{To upsample a 128 size latent vector to a 256*128*1 sample in Generator and to downsample a 256*128*1 sample to a size 1 output in Discriminator, we use Resnet architecture from the BigGAN paper \cite{Andrew_biggan}}. Table \ref{tab:tab1} and \ref{tab:tab2} show the layers of the architecture. Generator and Discriminator architectures are shown in Table \ref{tab:tab3} and \ref{tab:tab4}, respectively. 



We use a learning rate $0.00005$ and $0.0002$ for the Generator and the Discriminator, respectively. We set the number of channels (ch) to 16 to minimise the computational expenses as the higher number of channels such as 64 and 32 only offer negligible improvements.

\begin{table}[h!]
\centering
\caption{Architecture of the ResBlock generator with upsampling for the supervised BigGAN.}
\begin{tabular}{|l|l|l|l|}\hline
\textbf{\begin{tabular}[c]{@{}l@{}}Layer\\ Name\end{tabular}} & \textbf{\begin{tabular}[c]{@{}l@{}}Kernal\\ Size\end{tabular}} & \textbf{RS} & \textbf{\begin{tabular}[c]{@{}l@{}}Output\\ Size\end{tabular}} \\ \hline
Shortcut & {[}1,1,1{]} & U & 2h $\times$ 2w $\times$ c\_\{o\} \\ \hline
cBN, ReLU & - & - & h $\times$ w $\times$ c\_\{i\} \\ \hline
Convolution & {[}3,3,1{]} & U & 2h $\times$ 2w $\times$ c\_\{o\} \\ \hline
cBN, ReLU & - & - & 2h $\times$ 2w $\times$ c\_\{o\} \\ \hline
Convolution & {[}3,3,1{]} & U & 2h $\times$ 2w $\times$ c\_\{o\} \\ \hline
Addition & - & - & 2h $\times$ 2w $\times$ c\_\{o\} \\ \hline

\end{tabular}
\label{tab:tab1}
\end{table}

\begin{table}[h!]
\centering
\caption{Architecture of the ResBlock discriminator with downsampling for the supervised BigGAN.}
\begin{tabular}{|l|l|l|l|}
\hline
\textbf{\begin{tabular}[c]{@{}l@{}}Layer\\ Name\end{tabular}} & \textbf{\begin{tabular}[c]{@{}l@{}}Kernal\\ Size\end{tabular}} & \textbf{RS} & \textbf{\begin{tabular}[c]{@{}l@{}}Output\\ Size\end{tabular}} \\ \hline
Shortcut & {[}1,1,1{]} & D & h/2 $\times$ w/2 $\times$ c\_\{o\} \\ \hline
ReLU & - & - & h $\times$ w $\times$ c\_\{i\} \\ \hline
Convolution & {[}3,3,1{]} & - & h $\times$ w $\times$ c\_\{o\} \\ \hline
ReLU & - & - & h $\times$ w $\times$ c\_\{o\} \\ \hline
Convolution & {[}3,3,1{]} & D & h/2 $\times$ w/2 $\times$ c\_\{o\} \\ \hline
Addition & - & - & h/2 $\times$ w/2 $\times$ c\_\{o\} \\ \hline

\end{tabular}
\label{tab:tab2}
\end{table}


\begin{table}[h!]
\centering
\caption{Architecture of the generator for the supervised BigGAN.}
\begin{tabular}{|l|l|l|l|}
\hline
\textbf{\begin{tabular}[c]{@{}l@{}}Layer\\ Name\end{tabular}} & \textbf{RS} & \textbf{SN} & \textbf{\begin{tabular}[c]{@{}l@{}}Output\\ Size\end{tabular}} \\ \hline
Input z & - & - & 128 \\ \hline
Dense & - & - & 4 $\times$ 2 $\times$ 16. ch \\ \hline
ResBlock & U & SN & 8 $\times$ 4 $\times$ 16. ch \\ \hline
ResBlock & U & SN & 16 $\times$ 8 $\times$ 16. ch \\ \hline
ResBlock & U & SN & 32 $\times$ 16 $\times$ 16. ch \\ \hline
ResBlock & U & SN & 64 $\times$ 32 $\times$ 16. ch \\ \hline
ResBlock & U & SN & 128 $\times$ 64 $\times$ 16. ch \\ \hline
Non-local block & - & - & 128 $\times$ 64 $\times$ 16. ch \\ \hline
ResBlock & U & SN & 256 $\times$ 128 $\times$ 1. ch \\ \hline
BN, ReLU & - & - & 256 $\times$ 128 $\times$ 1 \\ \hline
Conv {[}3, 3, 1{]} & - & - & 256 $\times$ 128 $\times$ 1 \\ \hline
Tanh & - & - & 256 $\times$ 128 $\times$ 1 \\ \hline

\end{tabular}
\label{tab:tab3}
\end{table}

\begin{table}[h!]
\centering
\caption{Architecture of the discriminator for the supervised BigGAN.}
\begin{tabular}{|l|l|l|}
\hline
\textbf{\begin{tabular}[c]{@{}l@{}}Layer\\ Name\end{tabular}} & \textbf{RS} & \textbf{\begin{tabular}[c]{@{}l@{}}Output\\ Size\end{tabular}} \\ \hline
\begin{tabular}[c]{@{}l@{}}Input \\ Spectrogram\end{tabular} & - & 256 $\times$ 128 $\times$ 1 \\ \hline
ResBlock & D & 128 $\times$ 64 $\times$ 1. ch \\ \hline
Non-local block & - & 128 $\times$ 64 $\times$ 1. ch \\ \hline
ResBlock & - & 64 $\times$ 32 $\times$ 1. ch \\ \hline
ResBlock & D & 32 $\times$ 16 $\times$ 2. ch \\ \hline
ResBlock & D & 16 $\times$ 8 $\times$ 4. ch \\ \hline
ResBlock & D & 8 $\times$ 4 $\times$ 8. ch \\ \hline
ResBlock & D & 4 $\times$ 2 $\times$ 16. ch \\ \hline
\begin{tabular}[c]{@{}l@{}}ResBlock \\ (No Shortcut)\end{tabular} & - & 4 $\times$ 2 $\times$ 16. ch \\ \hline
ReLU & - & 4 $\times$ 2 $\times$ 16. ch \\ \hline
Global sum pooling & - & 1 $\times$ 1 $\times$ 16. ch \\ \hline
Sum(embed(y)·h)+(dense $\rightarrow$ 1) & - & 1 \\ \hline

\end{tabular}
\label{tab:tab4}
\end{table}

\subsection{Unsupervised BigGAN}

The architectures of upsampling and downsampling layers are given in table \ref{tab:tab5} and \ref{tab:tab6}, respectively. Here only the Batch Normalisation layers are different when compared to supervised BigGAN. Generator and Discriminator architectures are shown in the Table \ref{tab:tab7} and \ref{tab:tab8}, respectively. We use learning rate $0.00005$ and $0.0002$ for the Generator and the Discriminator, respectively. Again we set the number of channels to 16. 

\begin{table}[h!]
\centering
\caption{Architecture of the ResBlock generator with upsampling for the unsupervised BigGAN.}

\begin{tabular}{|l|l|l|l|}
\hline
\textbf{\begin{tabular}[c]{@{}l@{}}Layer\\ Name\end{tabular}} & \textbf{\begin{tabular}[c]{@{}l@{}}Kernal\\ Size\end{tabular}} & \textbf{RS} & \textbf{\begin{tabular}[c]{@{}l@{}}Output\\ Size\end{tabular}} \\ \hline
Shortcut & {[}1,1,1{]} & U & 2h $\times$ 2w $\times$ c\_\{o\} \\ \hline
BN, ReLU & - & - & h $\times$ w $\times$ c\_\{i\} \\ \hline
Convolution & {[}3,3,1{]} & U & 2h $\times$ 2w $\times$ c\_\{o\} \\ \hline
BN, ReLU & - & - & 2h $\times$ 2w $\times$ c\_\{o\} \\ \hline
Convolution & {[}3,3,1{]} & U & 2h $\times$ 2w $\times$ c\_\{o\} \\ \hline
Addition & - & - & 2h $\times$ 2w $\times$ c\_\{o\} \\ \hline

\end{tabular}
\label{tab:tab5}
\end{table}

\begin{table}[h!]
\centering
\caption{Architecture of the ResBlock discriminator with downsampling for the unsupervised BigGAN.}
\begin{tabular}{|l|l|l|l|}
\hline
\textbf{\begin{tabular}[c]{@{}l@{}}Layer\\ Name\end{tabular}} & \textbf{\begin{tabular}[c]{@{}l@{}}Kernal\\ Size\end{tabular}} & \textbf{RS} & \textbf{\begin{tabular}[c]{@{}l@{}}Output\\ Size\end{tabular}} \\ \hline
Shortcut & {[}1,1,1{]} & D & h/2 $\times$ w/2 $\times$ c\_\{o\} \\ \hline
ReLU & - & - & h $\times$ w $\times$ c\_\{i\} \\ \hline
Convolution & {[}3,3,1{]} & - & h $\times$ w $\times$ c\_\{o\} \\ \hline
ReLU & - & - & h $\times$ w $\times$ c\_\{o\} \\ \hline
Convolution & {[}3,3,1{]} & D & h/2 $\times$ w/2 $\times$ c\_\{o\} \\ \hline
Addition & - & - & h/2 $\times$ w/2 $\times$ c\_\{o\} \\ \hline

\end{tabular}
\label{tab:tab6}
\end{table}


\begin{table}[h!]
\centering
\caption{Architecture of the generator for the unsupervised BigGAN.}
\begin{tabular}{|l|l|l|l|}
\hline
\textbf{\begin{tabular}[c]{@{}l@{}}Layer\\ Name\end{tabular}} & \textbf{RS} & \textbf{SN} & \textbf{\begin{tabular}[c]{@{}l@{}}Output\\ Size\end{tabular}} \\ \hline
Input z & - & - & 128 \\ \hline
Dense & - & - & 4 $\times$ 2 $\times$ 16. ch \\ \hline
ResBlock & U & SN & 8 $\times$ 4 $\times$ 16. ch \\ \hline
ResBlock & U & SN & 16 $\times$ 8 $\times$ 16. ch \\ \hline
ResBlock & U & SN & 32 $\times$ 16 $\times$ 16. ch \\ \hline
ResBlock & U & SN & 64 $\times$ 32 $\times$ 16. ch \\ \hline
ResBlock & U & SN & 128 $\times$ 64 $\times$ 16. ch \\ \hline
Non-local block & - & - & 128 $\times$ 64 $\times$ 16. ch \\ \hline
ResBlock & U & SN & 256 $\times$ 128 $\times$ 1. ch \\ \hline
BN, ReLU & - & - & 256 $\times$ 128 $\times$ 1 \\ \hline
Conv {[}3, 3, 1{]} & - & - & 256 $\times$ 128 $\times$ 1 \\ \hline
Tanh & - & - & 256 $\times$ 128 $\times$ 1 \\ \hline

\end{tabular}
\label{tab:tab7}
\end{table}

\begin{table}[h!]
\centering
\caption{Architecture of the discriminator for the unsupervised BigGAN.}
\begin{tabular}{|l|l|l|}
\hline
\textbf{\begin{tabular}[c]{@{}l@{}}Layer\\ Name\end{tabular}} & \textbf{RS} & \textbf{\begin{tabular}[c]{@{}l@{}}Output\\ Size\end{tabular}} \\ \hline
\begin{tabular}[c]{@{}l@{}}Input \\ Spectrogram\end{tabular} & - & 256 $\times$ 128 $\times$ 1 \\ \hline
ResBlock & D & 128 $\times$ 64 $\times$ 1. ch \\ \hline
Non-local block & - & 128 $\times$ 64 $\times$ 1. ch \\ \hline
ResBlock & - & 64 $\times$ 32 $\times$ 1. ch \\ \hline
ResBlock & D & 32 $\times$ 16 $\times$ 2. ch \\ \hline
ResBlock & D & 16 $\times$ 8 $\times$ 4. ch \\ \hline
ResBlock & D & 8 $\times$ 4 $\times$ 8. ch \\ \hline
ResBlock & D & 4 $\times$ 2 $\times$ 16. ch \\ \hline
\begin{tabular}[c]{@{}l@{}}ResBlock \\ (No Shortcut)\end{tabular} & - & 4 $\times$ 2 $\times$ 16. ch \\ \hline
ReLU & - & 4 $\times$ 2 $\times$ 16. ch \\ \hline
Global sum pooling & - & 1 $\times$ 1 $\times$ 16. ch \\ \hline
Dense & - & 1 \\ \hline

\end{tabular}
\label{tab:tab8}
\end{table}

\subsection{GGAN}
In the GGAN model, the downsampling and upsampling layers are the same as those shown in table \ref{tab:tab5} and \ref{tab:tab6}, respectively.
For Encoder, we concatenate the input $z$ and $c$, then use a simple multi layer neural network. The Encoder architecture is shown in table \ref{tab:tab9}. Classifier, $C_{e}$ and $C_{x}$ share the same architecture, and it is shown in table \ref{tab:tab10}. We use the same configuration of the Generator as of the unsupervised GAN shown in table \ref{tab:tab7}. Architectures of the Feature extraction part ($D$) and the discrimination part ($D^{'}$) of the first Discriminator are given in  Table \ref{tab:tab12} and \ref{tab:tab13}, respectively. Also, for the second Discriminator, we use the same Discriminator architecture as that of the unsupervised BigGAN given in table \ref{tab:tab8}. Architecture of the Feature Extractor network is given in the table \ref{tab:tab15}. For the Discriminator $D_{f}$, we concatenate the features after the resblocks and before the bottleneck dense layer. The architecture is given in \ref{tab:tab17}. For all the networks except discriminators, we use the learning rate $0.0005$ and for the discriminators, we use the learning rate $0.0002$. We again set the number of channels to $16$. We use a batch size of 64 with two Nvidia p100 GPUs. The average run time is 33.12 hours.

\begin{table}[h!]
\centering
\caption{Architecture of the Encoder for the GGAN.}
\begin{tabular}{|l|l|}
\hline
\textbf{\begin{tabular}[c]{@{}l@{}}Layer\\ Name\end{tabular}} & \textbf{\begin{tabular}[c]{@{}l@{}}Output\\ Size\end{tabular}} \\ \hline
Input z, Input c & 128 + 10 = 138 \\ \hline
Dense & 128 \\ \hline
ReLU & 128 \\ \hline
Dense & 128 \\ \hline
ReLU & 128 \\ \hline
Dense & 128 \\ \hline

\end{tabular}
\label{tab:tab9}
\end{table}

\begin{table}[h!]
\centering
\caption{Architecture of the Classifier $C_{e}$ and $C_{x}$ for the GGAN.}
\begin{tabular}{|l|l|}
\hline
\textbf{\begin{tabular}[c]{@{}l@{}}Layer\\ Name\end{tabular}} & \textbf{\begin{tabular}[c]{@{}l@{}}Output\\ Size\end{tabular}} \\ \hline
Input & 128 \\ \hline
Dense & 128 \\ \hline
ReLU & 128 \\ \hline
Dense & 128 \\ \hline
ReLU & 128 \\ \hline
Dense & 10 \\ \hline

\end{tabular}
\label{tab:tab10}
\end{table}

\begin{table}[h!]
\centering
\caption{Architecture of the feature extraction part ($D$) of the first discriminator for the GGAN.}
\begin{tabular}{|l|l|l|}
\hline
\textbf{\begin{tabular}[c]{@{}l@{}}Layer\\ Name\end{tabular}} & \textbf{RS} & \textbf{\begin{tabular}[c]{@{}l@{}}Output\\ Size\end{tabular}} \\ \hline
\begin{tabular}[c]{@{}l@{}}Input \\ Spectrogram\end{tabular} & - & 256 $\times$ 128 $\times$ 1 \\ \hline
ResBlock & D & 128 $\times$ 64 $\times$ 1. ch \\ \hline
Non-local block & - & 128 $\times$ 64 $\times$ 1. ch \\ \hline
ResBlock & - & 64 $\times$ 32 $\times$ 1. ch \\ \hline
ResBlock & D & 32 $\times$ 16 $\times$ 2. ch \\ \hline
ResBlock & D & 16 $\times$ 8 $\times$ 4. ch \\ \hline
ResBlock & D & 8 $\times$ 4 $\times$ 8. ch \\ \hline
ResBlock & D & 4 $\times$ 2 $\times$ 16. ch \\ \hline
\begin{tabular}[c]{@{}l@{}}ResBlock \\ (No Shortcut)\end{tabular} & - & 4 $\times$ 2 $\times$ 16. ch \\ \hline
ReLU & - & 4 $\times$ 2 $\times$ 16. ch \\ \hline
Global sum pooling & - & 1 $\times$ 1 $\times$ 16. ch \\ \hline
Flatten & - & 16. ch \\ \hline

\end{tabular}
\label{tab:tab12}
\end{table}

\begin{table}[h!]
\centering
\caption{Architecture of the discrimination part ($D^{'}$) of the first discriminator for the GGAN.}
\begin{tabular}{|l|l|}
\hline
\textbf{\begin{tabular}[c]{@{}l@{}}Layer\\ Name\end{tabular}} & \textbf{\begin{tabular}[c]{@{}l@{}}Output\\ Size\end{tabular}} \\ \hline
Input z & 128 \\ \hline
Dense & 128 \\ \hline
ReLU & 128 \\ \hline
Dense & 128 \\ \hline
ReLU & 128 \\ \hline
Dense & 1 \\ \hline

\end{tabular}
\label{tab:tab13}
\end{table}

\begin{table}[h!]
\centering
\caption{Architecture of the Feature Extractor for the GGAN.}
\begin{tabular}{|l|l|}
\hline
\textbf{\begin{tabular}[c]{@{}l@{}}Layer\\ Name\end{tabular}} & \textbf{\begin{tabular}[c]{@{}l@{}}Output\\ Size\end{tabular}} \\ \hline
Input Feature & 256 \\ \hline
Dense & 256 \\ \hline
ReLU & 256 \\ \hline
Dense & 256 \\ \hline
ReLU & 256 \\ \hline
Dense & 128 \\ \hline

\end{tabular}
\label{tab:tab15}
\end{table}

\begin{table}[ht!]
\centering
\caption{Architecture of the Discriminator, $D_{f}$ for the GGAN.}
\begin{tabular}{|l|l|l|}
\hline
\textbf{\begin{tabular}[c]{@{}l@{}}Layer\\ Name\end{tabular}} & \textbf{RS} & \textbf{\begin{tabular}[c]{@{}l@{}}Output\\ Size\end{tabular}} \\ \hline
\begin{tabular}[c]{@{}l@{}}Input \\ Spectrogram\end{tabular} & - & 256 $\times$ 128 $\times$ 1 \\ \hline
ResBlock & D & 128 $\times$ 64 $\times$ 1. ch \\ \hline
Non-local block & - & 128 $\times$ 64 $\times$ 1. ch \\ \hline
ResBlock & - & 64 $\times$ 32 $\times$ 1. ch \\ \hline
ResBlock & D & 32 $\times$ 16 $\times$ 2. ch \\ \hline
ResBlock & D & 16 $\times$ 8 $\times$ 4. ch \\ \hline
ResBlock & D & 8 $\times$ 4 $\times$ 8. ch \\ \hline
ResBlock & D & 4 $\times$ 2 $\times$ 16. ch \\ \hline
\begin{tabular}[c]{@{}l@{}}ResBlock \\ (No Shortcut)\end{tabular} & - & 4 $\times$ 2 $\times$ 16. ch \\ \hline
ReLU & - & 4 $\times$ 2 $\times$ 16. ch \\ \hline
Global sum pooling & - & 1 $\times$ 1 $\times$ 16. ch \\ \hline
Concat with input feature  & - & 256+128=384 \\ \hline
Dense & - & 128 \\ \hline
ReLU & - & 128 \\ \hline
Dense & - & 1 \\ \hline

\end{tabular}
\label{tab:tab17}
\end{table}





\ifCLASSOPTIONcaptionsoff
  \newpage
\fi

\begin{IEEEbiography}[{\includegraphics[width=1.1in,height=1.1in,clip]{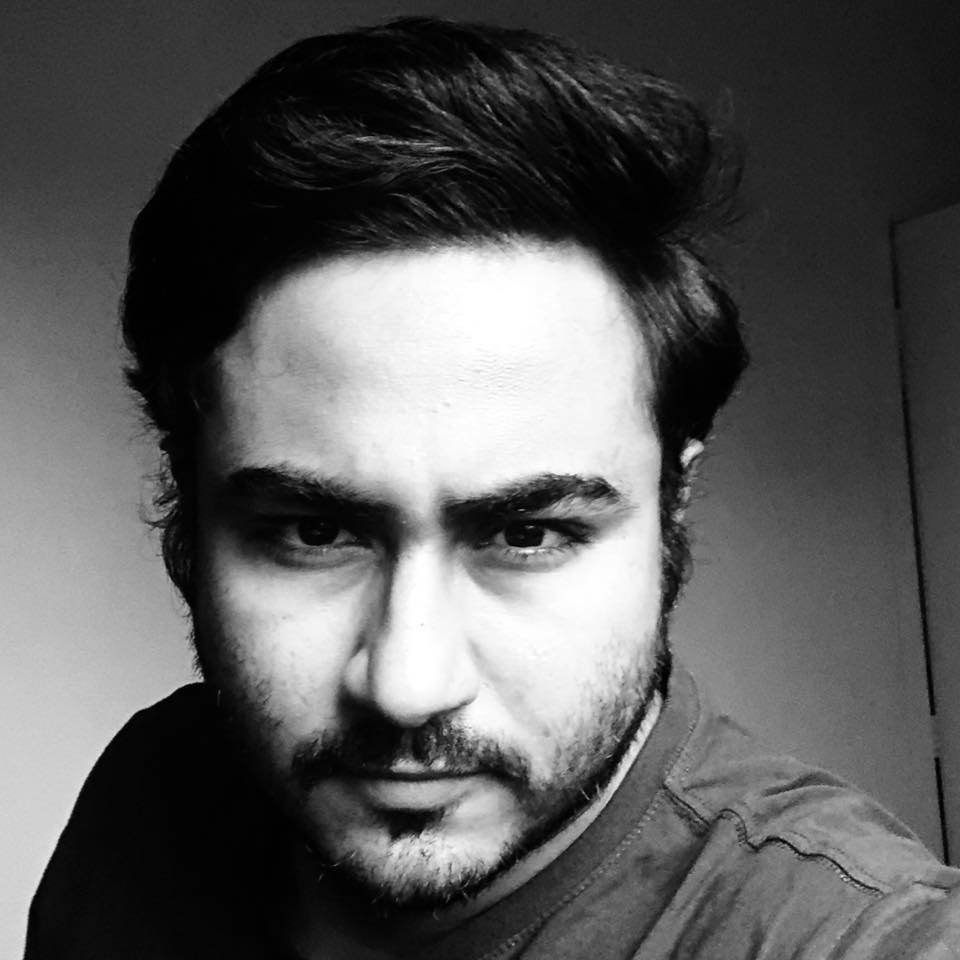}}]{Kazi Nazmul Haque}
is a PhD student at University of Southern Queensland, Australia. He has been working professionally in the field of machine learning for more than five years. Kazi's research work focuses on building machine learning models to solve diverse real-world problems. The current focus of his research work is unsupervised representation learning for the audio data. He has completed his Master in Information Technology from Jahangirnagar University, Bangladesh. 
\end{IEEEbiography}

\begin{IEEEbiography}[{\includegraphics[width=1.1in,height=1.1in,clip]{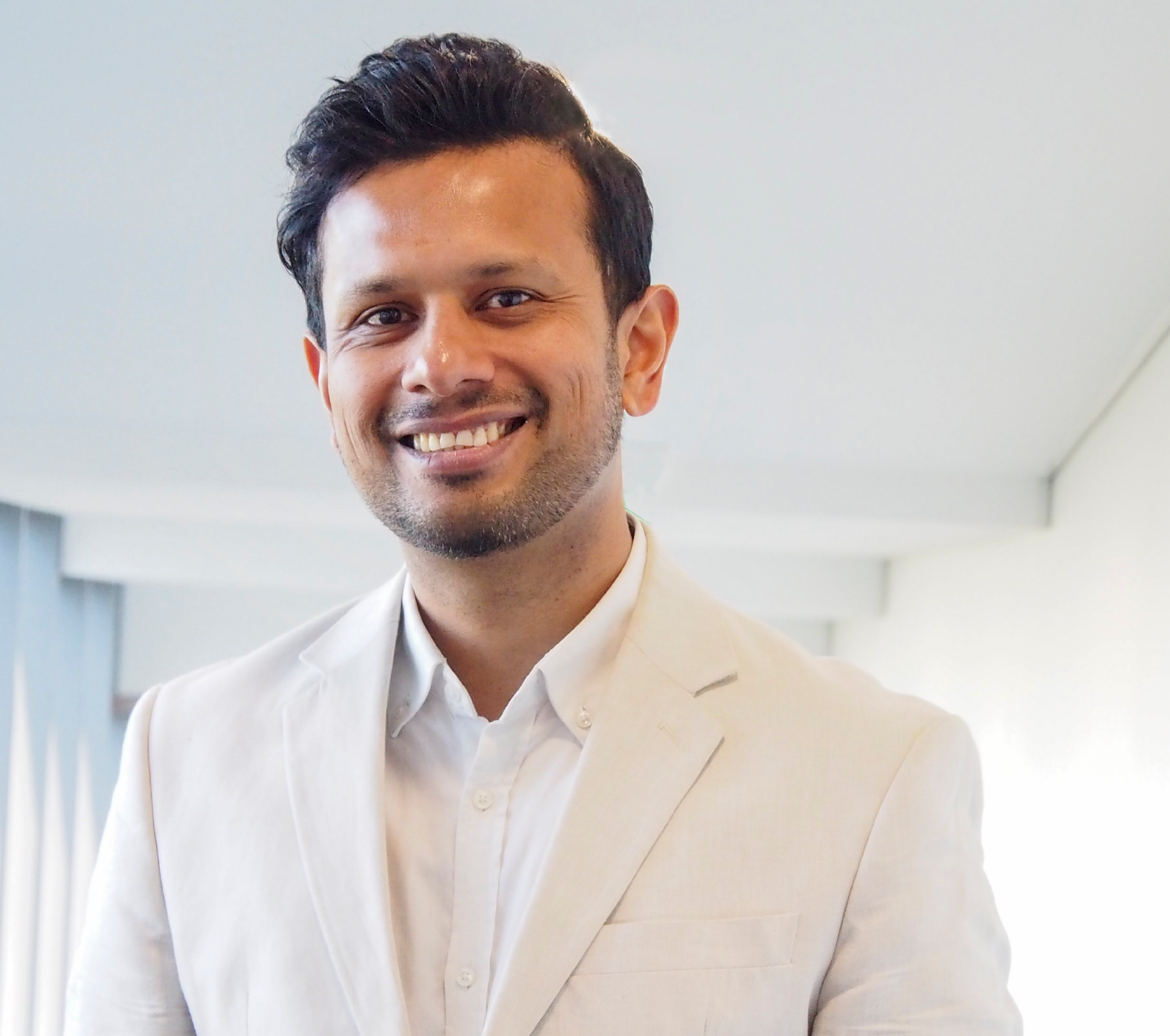}}]{Rajib Rana}
is an experimental computer scientist, Advance Queensland Research Fellow and a Senior Lecturer in the University of Southern Queensland. He is also the Director of IoT Health research program at the University of Southern Queensland. He is the recipient of the prestigious Young Tall Poppy QLD Award 2018 as one of Queensland’s most outstanding scientists for achievements in the area of scientific research and communication. Rana's research work aims to capitalise on advancements in technology along with sophisticated information and data processing to better understand disease progression in chronic health conditions and develop predictive algorithms for chronic diseases, such as mental illness and cancer. His current research focus is on Unsupervised Representation Learning. He received his B.Sc. degree in Computer Science and Engineering from Khulna University, Bangladesh with Prime Minister and President’s Gold medal for outstanding achievements and Ph.D. in Computer Science and Engineering from the University of New South Wales, Sydney, Australia in 2011. He received his postdoctoral training at Autonomous Systems Laboratory, CSIRO before joining the University of Southern Queensland as Faculty in 2015.
\end{IEEEbiography}

\begin{IEEEbiography}[{\includegraphics[width=1.1in,height=1.1in,clip]{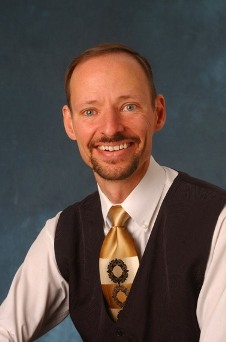}}]{John H. L. Hansen (S’81–M’82–SM’93–F’07)}
 received the Ph.D. and M.S. degrees in electrical
engineering from Georgia Institute of Technology,
Atlanta, GA, USA, in 1988 and 1983, respectively,
and the B.S.E.E. degree from Rutgers University,
College of Engineering, New Brunswick, NJ, USA,
in 1982. He received the Honorary degree Doctor
Technices Honoris Causa from Aalborg University,
Aalborg, Denmark, in April 2016 in recognition of
the contributions to speech signal processing and
speech/language/hearing sciences. He is ISCA President for 2017–2019 term. He joined the University of Texas at Dallas (UT
Dallas), Erik Jonsson School of Engineering and Computer Science, Richardson, TX, USA, in 2005, where he is currently a Jonsson School Associate Dean
for Research, as well as a Professor of electrical and computer engineering, the
Distinguished University Chair in telecommunications engineering, and a joint
appointment as a Professor with the School of Behavioral and Brain Sciences
(Speech \& Hearing). He was the Department Head of electrical engineering from
August 2005–December 2012, overseeing a +4x increase in research expenditures (4.5 M–22.3 M) with a 20
additional T/TT faculty, growing UT Dallas to the 8th largest EE program from
ASEE rankings in terms of degrees awarded. At UT Dallas, he established the
Center for Robust Speech Systems (CRSS). He was the Department Chairman
and a Professor with the Department of Speech, Language and Hearing Sciences
(SLHS), and a Professor in electrical and computer engineering, University of
Colorado Boulder, Boulder, CO, USA (1998–2005), where he co-founded and
was the Associate Director of the Center for Spoken Language Research. In
1988, he established the Robust Speech Processing Laboratory and continues to
direct research activities in CRSS, UT Dallas. He is author/coauthor of 661 journal and conference papers including 12 textbooks in the field of speech processing and language technology, signal processing for vehicle systems, coauthor of
textbook Discrete-Time Processing of Speech Signals, (IEEE Press, 2000), coeditor of the DSP for InVehicle and Mobile Systems (Springer, 2004), Advances
for In-Vehicle and Mobile Systems: Challenges for International Standards
(Springer, 2006), InVehicle Corpus and Signal Processing for Driver Behavior (Springer, 2008), and lead author of the report The Impact of Speech Under
Stress on Military Speech Technology, (NATO RTO-TR-10, 2000). His research
interests include digital speech processing, analysis and modeling of speech and
speaker traits, speech enhancement, feature estimation in noise, robust speech
recognition with emphasis on spoken document retrieval, and in-vehicle interactive systems for hands-free human-computer interaction. He has been named
IEEE Fellow (2007) for contributions in Robust Speech Recognition in Stress
and Noise, International Speech Communication Association (ISCA) Fellow
(2010) for contributions on research for speech processing of signals under adverse conditions. He was the recipient of The Acoustical Society of Americas
25 Year Award (2010) in recognition of his service, contributions, and membership to the Acoustical Society of America. He is currently the ISCA President
(2017–2019) and a member of the ISCA Board, having previously served as
the Vice-President (2015–2017). He also was selected and is the Vice-Chair
on U.S. Office of Scientific Advisory Committees (OSAC), Washington, DC,
USA, for OSAC-Speaker in the voice forensics domain (2015–2017). He was
the IEEE Technical Committee (TC) Chair and a Member of the IEEE Signal
Processing Society: Speech-Language Processing Technical Committee (SLTC)
(2005–2008; 2010–2014; elected IEEE SLTC Chairman for 2011-2013, PastChair for 2014), and elected as a ISCA Distinguished Lecturer (2011–2012).
He was a member of IEEE Signal Processing Society Educational Technical
Committee (2005–2008; 2008–2010); a Technical Advisor to the U.S. Delegate
for NATO (IST/TG-01); IEEE Signal Processing Society Distinguished Lecturer (2005/2006), an Associate Editor of the IEEE TRANSACTIONS ON SPEECH
AND AUDIO PROCESSING (1992–1999), an Associate Editor of the IEEE SIGNAL PROCESSING LETTERS (1998–2000), Editorial Board Member of the IEEE
Signal Processing Magazine (2001–2003); and a Guest Editor (October 1994)
for special issue on Robust Speech Recognition of the IEEE TRANSACTIONS
ON SPEECH AND AUDIO PROCESSING. He has served on Speech Communications Technical Committee for Acoustical Society of America (2000-2003),
and previously on ISCA Advisory Council. He has supervised 82 Ph.D./M.S.
thesis candidates (45 Ph.D., 37 M.S./M.A.), received The 2005 University of
Colorado Teacher Recognition Award as voted on by the student body. He also
organized and was s the General Chair for ISCA Interspeech- 2002, September
16–20, 2002, Co-Organizer and Technical Program Chair for IEEE ICASSP2010, Dallas, TX, USA, March 15–19, 2010, and Cochair and Organizer for
IEEE SLT-2014, December 7–10, 2014, in Lake Tahoe, NV, USA.
 \end{IEEEbiography}

\begin{IEEEbiography}[{\includegraphics[width=1in,height=1.25in,clip,keepaspectratio]{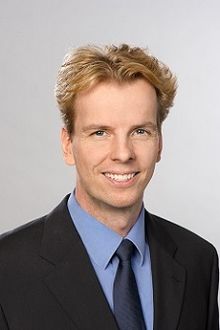}}]{Bj\"{o}rn W.\ Schuller  (M'05-SM'15-F'18)} received his diploma in 1999, his doctoral degree for his study
on Automatic Speech and Emotion Recognition in 2006, and his habilitation and Adjunct Teaching Professorship in the subject area of Signal Processing and Machine Intelligence in 2012, all
in electrical engineering and information technology from TUM in Munich/Germany. He is Professor of Artificial Intelligence in the Department of Computing at the Imperial College London/UK, where he heads GLAM –- the Group on Language, Audio \& Music, Full Professor and head of the ZD.B Chair of Embedded Intelligence for Health Care and Wellbeing at the University of Augsburg/Germany, and CEO of audEERING. He was previously full
professor and head of the Chair of Complex and Intelligent Systems at the University of Passau/Germany. Professor Schuller is Fellow of the IEEE, Golden Core Member of the IEEE Computer Society, Senior Member of the ACM, President-emeritus of the Association for the Advancement of Affective Computing (AAAC), and was elected member of the IEEE Speech and Language Processing Technical Committee. He (co-)authored 5 books and more than 800 publications in peer-reviewed books, journals, and conference proceedings leading to more than overall 25\,000 citations (h-index = 73). Schuller is general chair of ACII 2019, co-Program Chair of Interspeech 2019 and ICMI 2019, repeated Area Chair of ICASSP, and former Editor in Chief of the IEEE Transactions on Affective Computing next to a multitude of further Associate and Guest Editor roles and functions in Technical and Organisational Committees.
\end{IEEEbiography}






\end{document}